\documentclass[doublecol]{epl2}

\usepackage[utf8]{inputenc}
\usepackage{graphicx}
\usepackage{dcolumn}
\usepackage{color}
\usepackage[pdftex]{hyperref}
\usepackage{amsmath}
\usepackage{amssymb}
\usepackage{cite}
\usepackage{ulem}

\hypersetup{
  bookmarks=true,         % show bookmarks bar?
  unicode=false,          % non-Latin characters in Acrobat?s bookmarks
  pdftoolbar=true,        % show Acrobat?s toolbar?
  pdfmenubar=true,        % show Acrobat?s menu?
  pdffitwindow=true,      % page fit to window when opened
  pdfnewwindow=true,      % links in new window
  colorlinks=false,       % false: boxed links; true: colored links
  linkcolor=red,          % color of internal links
  citecolor=blue,        % color of links to bibliography
  filecolor=magenta,      % color of file links
  urlcolor=blue           % color of external
}

\title{Parametric resonance and spin--charge separation in 1D fermionic systems}

\author{Christian D.\ Graf\footnote{These authors contributed equally to this
work.}\hspace{.15cm} \inst{1} \and
Guillaume Weick$^\ast$\inst{1,2}\and
Eros Mariani\inst{1,3}}

\shortauthor{C.\ D.\ Graf, G.\ Weick and E.\ Mariani}

\institute{
\inst{1}Dahlem Center for Complex Quantum Systems \& Fachbereich Physik, 
Freie Universit\"at Berlin, D-14195 Berlin, Germany\\
\inst{2}Institut de Physique et Chimie des Mat\'eriaux de Strasbourg, UMR 7504
CNRS-UdS, F-67034 Strasbourg, France\\
\inst{3}School of Physics, University of Exeter, Stocker Road, Exeter, EX4 4QL, United Kingdom
}

\date{\today}


\abstract{
We show that the periodic modulation of the Hamiltonian parameters for 1D 
correlated fermionic systems can be used to parametrically amplify their
bosonic collective modes. Treating the problem within the Luttinger liquid
picture, we show how charge and spin density waves with different momenta are 
simultaneously amplified. We discuss the implementation of our predictions for 
cold atoms in 1D modulated optical lattices, showing that the fermionic momentum distribution directly provides a clear signature of spin--charge separation.}

\begin{document}

\maketitle

%===========================================================================
%===========================================================================
%===========================================================================
%===========================================================================
\section{Introduction}
The swing is the best known example of a classical system 
showing parametric resonances \cite{landau}. The periodic modulation of the effective 
swing length induced by the motion of legs leads to the exponential amplification 
of the oscillations if the modulation frequency is chosen commensurately with 
the natural frequency of the swing.
Quantizing this classical problem as a harmonic oscillator with modulated
parabolic confinement leads to the appearance of an exponential divergence in
the time evolution of the bosonic rising and lowering operators. This effect is
particularly strong if the modulation is around twice the natural oscillator frequency. 

In a modulated system of many bosonic oscillators only those fulfilling the resonance condition
will be amplified, making parametric resonance a spectroscopic tool in many-body quantum systems.
These ideas acquired particular relevance since cold atoms in optical lattices
have been realized \cite{bloch08_RMP}. As the intensity of the lattice can be
fully controlled by the laser power one can study parametric modulations in
correlated quantum systems with current experimental tools \cite{stofe04_PRL}.
Similarly, the periodic modulation of the transverse confinement in
cigar-shaped Bose--Einstein condensates has been shown to induce the parametric
amplification of Faraday waves \cite{engel07_PRL}. From the theoretical point of
view, parametric amplification of Bogoliubov quasiparticles for bosonic clouds
in optical lattices have been already investigated in the past
\cite{tozzo05_PRA, goren07_PRA}. In these systems, the bosonic nature of
quasiparticles appears already when interactions are treated at mean-field
level, and allows for the amplification to occur. 

\begin{figure}[tbh]
\onefigure[width=.6\columnwidth]{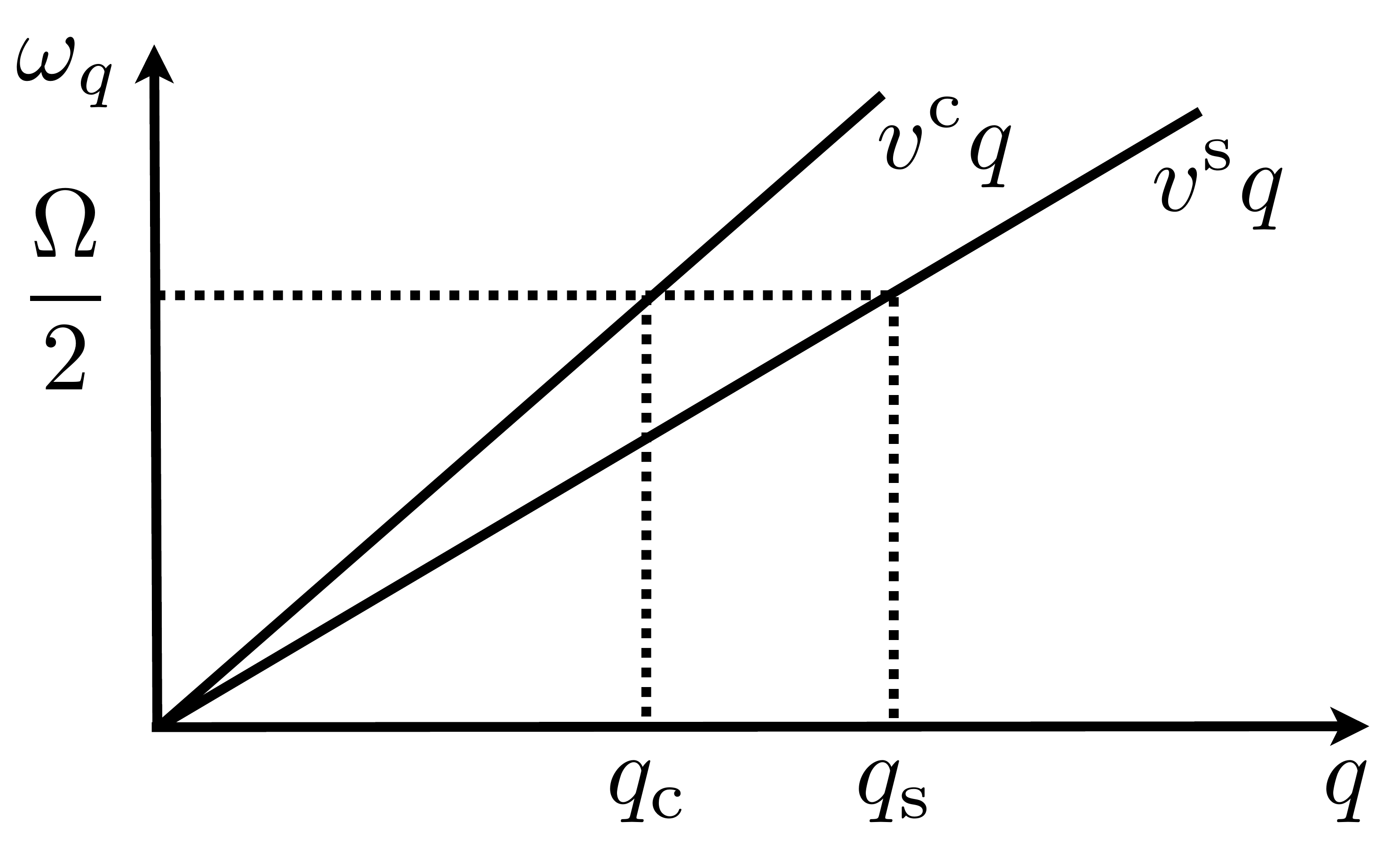}
\caption{\label{fig:dispersion}%
Sketch of the dispersions $\omega_q$ of the charge 
and spin density waves as a function of momentum $q$, with group velocities $v^\textrm{c}$ and
$v^\textrm{s}$, respectively. The pumping frequency $\Omega$ amplifies
collective spin and charge modes in the vicinity of the momenta $q_\textrm{c}$
and  $q_\textrm{s}$.}
\end{figure}

In this paper we analyze parametric resonances in many-body \textit{fermionic} systems,
starting with the very question whether the amplification can occur at all.
Indeed, in contrast to the bosonic case, 
in fermionic systems any
mean-field treatment of interactions, including the presence of broken
symmetries, preserves the fermionic nature of quasiparticles. The Pauli
principle thus blocks their amplification, as can be easily checked by direct
calculation \cite{goren07_PRA}. Then the question rises if bosonic collective
excitations of a fermionic many-body system can be subject to amplification by modulating a parameter in the microscopic Hamiltonian.
In order to address this fundamental question we need to treat correlations in a
fermionic system beyond mean-field level. In this work we thus confine our
investigation to one-dimensional (1D) correlated fermions within the Luttinger
liquid picture, in which interactions are treated exactly and the system is
naturally diagonalized in terms of collective bosonic spin and charge density
waves \cite{giuliani, giamarchi}. 
According to the Luttinger liquid theory these modes disperse with two different
group velocities, giving rise to the so-called spin--charge separation (see
fig.~\ref{fig:dispersion}). This
fundamental issue in condensed matter physics has been detected in transport
experiments on quantum wires \cite{ausla05_Science} and by angle-resolved
photoemission spectroscopy of 1D SrCuO$_2$ \cite{kim06_NaturePhysics} only recently. 
Due to their great tunability, cold atomic gases in optical lattices are also
promising candidates for the experimental detection of spin--charge
separation in 1D systems, as shown by several theoretical proposals
\cite{recat03_PRL, kecke05_PRL, kolla05_PRL, polin07_PRL, klein08_PRA, mathe08_PRL, kagan09_PRA}.

Here we show how a spatially homogeneous
time-periodic modulation of the intensity of the optical lattice indeed leads to
the amplification of charge and spin density waves of a 1D correlated cloud of
ultracold fermionic atoms. If $\Omega$ is the modulation frequency, the charge
and spin waves with energy $v^\textrm{c}q_\textrm{c}=v^\textrm{s}q_\textrm{s}
=\Omega /2$ will be amplified (see fig.~\ref{fig:dispersion}). 
Due to the different group velocities $v^\textrm{c}$ and $v^\textrm{s}$ for the
charge and spin channels, respectively,
the resonant condition amplifies different wavenumbers 
for the two branches, $q_\textrm{c}$ and $q_\textrm{s}$. On top of showing the feasibility of 
parametric amplification in correlated fermionic systems, we also propose this 
technique as a tool to systematically investigate the spin--charge separation in experiments. 
Indeed, we discuss the effect of the amplification above on the fermionic momentum distribution and 
show how the latter exhibits well defined shoulders directly related to the wavenumbers 
$q_\textrm{c}$ and $q_\textrm{s}$. 
These structures are particularly evident after not too long modulation times.
As the momentum distribution is the standard quantity 
measured in time-of-flight experiments on cold atomic clouds \cite{bloch08_RMP}, 
our analysis has direct implications on investigations of correlated fermionic
systems with current experimental tools.

%===========================================================================
%===========================================================================
%===========================================================================
%===========================================================================
\section{Model}
We consider interacting fermionic atoms with (pseudo-)spin $1/2$ 
confined into 1D cigars (as,
\textit{e.g.}, realized in optical lattices \cite{bloch08_RMP}), 
with a further periodic potential along the 1D axis. Despite this additional
potential, we assume the atoms to be
in their metallic phase, as opposed to the recently
investigated 
Mott-insulator phase \cite{jorde08_Nature, schne08_Science,
masse09_PRL}. In order to obtain analytical results, we disregard trapping and finite
size effects, as they will not qualitatively modify our results.
The parametric excitation of the system
is achieved by periodically modulating the intensity of the optical lattice 
along the 1D system, thereby shrinking the Wannier wavefunctions associated to each 
lattice site. 
This has the two-fold 
effect of modulating the hopping rate between neighboring sites (\textit{i.e.},
the kinetic energy) as well as the on-site repulsion for multiple occupancy. As the former is exponentially sensitive to the wavefunction overlap it is the dominant parametric modulation term in the problem and we will focus on this for simplicity. 
It should be however noticed that the inclusion of the smaller parametric modulation of the interactions would not 
significantly affect the consequences discussed in the following.

Once translated into the Bloch-band language, and if we focus on the low-energy physics of the problem, the modulation in the kinetic term yields an effectively time-dependent Fermi velocity for the atoms close to the Fermi level, $v_\textrm{F}(t)=v_\textrm{F}+\delta v_\textrm{F}(t)$. Here $\delta v_\textrm{F}(t)=\gamma v_\textrm{F}\sin{(\Omega t)}$,
where $v_\textrm{F}$ is the Fermi velocity at equilibrium (\textit{i.e.}, for times $t<0$
before the modulation starts), and $\Omega$ and $\gamma$ are the
frequency and intensity of the parametric modulation, respectively. 
Our focus on the low-energy sector of the many-body problem naturally suggests a
Luttinger liquid approach \cite{giuliani, giamarchi} to the correlated system, which allows for an exact treatment of interactions. We thus linearize the single-particle spectrum in the vicinity of the Fermi level for momenta $||k|-k_\textrm{F}|<\Lambda$, 
with $k_\textrm{F}$ the Fermi wavevector and $\Lambda$ an ultraviolet cutoff.
The time-dependent Hamiltonian describing the system of size $L$ 
therefore reads\footnote{In the remaining of the paper, we set
$\hbar=k_\textrm{B}=1$.}
\begin{align}
\label{eq:H(t)}
H(t)=v_\textrm{F}(t)\sum_{k \sigma\tau}f_{k\tau}
c_{k\sigma\tau}^\dagger c_{k\sigma\tau}
+\sum_{\substack{q\neq0\\ \tau\tau' \sigma\sigma'}}
\frac{V_{\tau \tau' ,q}^{\sigma\sigma'}}{2L}\rho^{\tau\sigma}_{-q}\rho^{\tau'\sigma'}_{q}
\end{align}
with $f_{k\tau}=\tau k-k_\textrm{F}$.
Here, $c_{k\sigma\tau}^\dagger$ ($c_{k\sigma\tau}$) creates (annihilates) a
$\tau$-moving fermion with momentum
$k$ and spin $\sigma=\uparrow,\downarrow$ ($\tau=1$ ($-1$) corresponds to right
(left) movers) while $\rho^{\tau\sigma}_{q}=\sum_k c_{k-q,\sigma ,\tau}^\dagger c_{k,\sigma ,\tau}$ is the corresponding fermionic density operator. 
In (\ref{eq:H(t)}) we keep the general form of the interaction between fermionic densities 
with generic spin and branch indices. Within the standard Luttinger liquid theory we 
have $V_{\tau \tau ,q}^{\sigma\sigma }=V^{\parallel}_{0}$, 
$V_{\tau \tau ,q}^{\sigma,-\sigma }=V^{\perp}_{0}$, 
$V_{\tau ,-\tau ,q}^{\sigma\sigma }=V^{\parallel}_{0}-V^{\parallel}_{2k_{{\rm F}}^{}}$, 
$V_{\tau ,-\tau ,q}^{\sigma, -\sigma }=V^{\perp}_{0}$, where $V^{\parallel /\perp }_{q}$ 
is the Fourier transform of the microscopic interaction between fermions with 
parallel/antiparallel spins. This allows to treat short range interactions (like
contact $s$-wave scattering for neutral fermions) where Pauli principle imposes 
$V^{\parallel}_{q}=0$, as well as finite range spin-invariant ones (\textit{e.g.}, 
between charged fermions) with $V^{\parallel}_{q}=V^{\perp}_{q}$. 
Eq.~\eqref{eq:H(t)} does not include backscattering between
particles with opposite spins as well as umklapp scattering as those terms are
usually negligible at equilibrium and away from half-filling \cite{giuliani, giamarchi}. 
The effect of the smaller periodic modulation of these interaction terms on the energy 
absorption has been considered in ref.~\cite{iucci06_PRA}.

Introducing the bosonic operators for charge and spin density fluctuations
$b_q^\textrm{c}=(\pi/L|q|)^{1/2}\sum_\tau\Theta(\tau q)
(\rho^{\tau\uparrow}_q+\rho^{\tau\downarrow}_q)$
and 
$b_q^\textrm{s}=(\pi/L|q|)^{1/2}\sum_\tau\Theta(\tau q)
(\rho^{\tau\uparrow}_q-\rho^{\tau\downarrow}_q)$,
respectively, 
eq.~\eqref{eq:H(t)} transforms into the separable Hamiltonian \cite{giuliani}
\begin{equation}
\label{eq:spin-charge}
H(t)=
\sum_{\substack{a=\textrm{c,s}\\ q\neq0}}|q|\left[A_q^a(t)
b_q^{a\dagger} b_q^a
+B_q^a\left(b_{q}^{a\dagger} b_{-q}^{a\dagger}
+b_{-q}^ab_{q}^a\right)\right],
\end{equation}
with $A_q^\textrm{c}(t)=v_\textrm{F}(t)+(V^{\parallel}_{0}+V^{\perp}_{0})/2\pi$,
$A_q^\textrm{s}(t)=v_\textrm{F}(t)+(V^{\parallel}_{0}-V^{\perp}_{0})/2\pi$, 
$B_q^\textrm{c}=(V^{\parallel}_{0}+V^{\perp}_{0}-V^{\parallel}_{2k_{{\rm F}}^{}})/4\pi$, and 
$B_q^\textrm{s}=(V^{\parallel}_{0}-V^{\perp}_{0}-V^{\parallel}_{2k_{{\rm F}}^{}})/4\pi$.

The time-independent part in \eqref{eq:spin-charge}
can be diagonalized by means of the Bogoliubov transformation
$b_q^{a}=\cosh{\varphi_q^a}\,\beta_q^{a}
+\sinh{\varphi_q^a}\,\beta_{-q}^{a\dagger}$ in terms of the bosonic fields $\beta_q^{a}$, 
such that 
\begin{align}
\label{eq:H_a}
H(t)=&\sum_{a, q\neq0}\left[
\left(\omega_q^a+\delta v_\textrm{F}(t)|q|\cosh(2\varphi_q^a)\right) 
\beta_q^{a\dagger}\beta_q^a\phantom{\frac 12}\right.\nonumber\\
&\left.+\frac{\delta v_\textrm{F}(t)}{2}|q|\sinh(2\varphi_q^a)
\left(
\beta_{q}^{a\dagger}\beta_{-q}^{a\dagger}
+\beta_{-q}^a\beta_{q}^a
\right)
\right]
\end{align}
with $\omega_q^a=v_q^a|q|$. The different charge and spin group velocities are 
$v_q^a=[{A_q^a(0)}^2-4{B_q^a}^2]^{1/2}$. 
For short range interactions, one can neglect the momentum dependence of the group
velocities such that $v_q^a\simeq v^a$, and the dispersion of the spin and
charge density waves is linear (see
fig.~\ref{fig:dispersion}). The coefficients of the Bogoliubov transformation read
$\sinh{\varphi_q^\textrm{c}}=
-[A_q^\textrm{c}(0)/2v_q^\textrm{c}-1/2]^{1/2}$,
$\cosh{\varphi_q^\textrm{c}}=
[A_q^\textrm{c}(0)/2v_q^\textrm{c}+1/2]^{1/2}$,
$\sinh{\varphi_q^\textrm{s}}=
[A_q^\textrm{s}(0)/2v_q^\textrm{s}-1/2]^{1/2}$,
and $\cosh{\varphi_q^\textrm{s}}=
[A_q^\textrm{s}(0)/2v_q^\textrm{s}+1/2]^{1/2}$.

The crucial point to notice at this level is that the parametric modulation
introduces a time-dependent anomalous term in the Hamiltonian \eqref{eq:H_a},
creating and annihilating pairs of bosonic charge and spin density waves. As the
modulation is homogeneous in space (\textit{i.e.}, at zero wavenumber), the new terms 
create or annihilate pairs of excitations with opposite wavenumber, as requested by 
momentum conservation \cite{goren07_PRA}. In addition, the induced anomalous terms 
are proportional to $\sinh(2\varphi_q^a)$ and thus correctly vanish in the 
non-interacting limit $V_{q}^{\parallel/\perp}=0$,
where fermionicity forbids parametric amplification.

%===========================================================================
%===========================================================================
%===========================================================================
%===========================================================================
\section{Parametric amplification}
From the Hamiltonian \eqref{eq:H_a} above we can now determine the time evolution of the operators $\beta_q^a$, showing that indeed parametric amplification of collective bosonic modes in a fermionic system is
possible.
With eq.~\eqref{eq:H_a}, the Heisenberg equation of motion
for the operator $\beta_q^a$ reads
$\dot\beta_q^a(t)=
-\textrm{i}[\omega_q^a+\delta v_\textrm{F}(t)|q|\cosh{(2\varphi_q^a)}]\beta_q^a(t)
-\textrm{i}\delta
v_\textrm{F}(t)|q|\sinh{(2\varphi_q^a)}\beta_{-q}^{a\dagger}(t)$.
Defining 
$\beta_q^a(t)=
\textrm{e}^{-\textrm{i}\int_0^t\textrm{d}s
[\omega_q^a+\delta v_\textrm{F}(s)|q|\cosh{(2\varphi_q^a)}]}\tilde\beta_q^a(t)$, 
assuming a weak parametric modulation ($\gamma\ll1$) 
and retaining only slow terms near the 
resonance (rotating wave approximation, \textit{i.e.}, for $\Omega$ in the vicinity of
$2\omega_q^a$), the equation of motion
and its adjoint transform
into \cite{tozzo05_PRA, goren07_PRA}
$\ddot{\tilde{\beta}}_q^a(t)+\textrm{i}(\Omega-2\omega_q^a)\dot{\tilde{\beta}}_q^a(t)
-{\xi_q^a}^2\tilde\beta_q^a(t)=0$, 
where $\xi_q^a=\gamma v_\textrm{F}|q|\sinh{(2\varphi_q^a)}/2$. Solving
this equation with the appropriate initial conditions $\tilde\beta_q^a(0)=\beta_q^a(0)$
and $\dot{\tilde{\beta}}_q^a(0)=\xi_q^a\beta_{-q}^{a\dagger}(0)$, we obtain 
\begin{equation}
\label{eq:beta}
\beta_q^a(t)=\sum_{\eta=\pm}\eta\,\textrm{e}^{\textrm{i}(\omega_{q\eta}^a-\omega_q^a)t}
\left[
\bar\omega_{q,-\eta}^a\beta_q^a(0)+\textrm{i}\,\bar\xi_q^a\beta_{-q}^{a\dagger}(0)
\right],
\end{equation}
where 
$\omega_{q\pm}^a=\omega_q^a-\Omega/2
\pm\sqrt{\left(\omega_q^a-\Omega/2\right)^2-{\xi_q^a}^2}$.
In \eqref{eq:beta}, we defined
$\bar\omega_{q\pm}^a=\omega_{q\pm}^a/(\omega_{q-}^a-\omega_{q+}^a)$ and 
$\bar\xi_{q}^a=\xi_{q}^a/(\omega_{q-}^a-\omega_{q+}^a)$. Thus, in a narrow ``resonant window" of
energy $|\omega_q^a-\Omega/2|<|\xi_q^a|$, the frequencies $\omega_{q\pm}$
acquire an imaginary part, leading to the exponential amplification of the corresponding bosonic modes. Outside this window, the modes evolve according to their coherent
dynamics and are therefore not amplified. Indeed, out of eq.~(\ref{eq:beta}), it is easy to verify that, on- and off-resonance, the evolution of the Bogoliubov operators is
\begin{subequations}
\label{eq:betaONOFF}
\begin{align}
\beta_q^a(t)\simeq
\textrm{e}^{-\textrm{i}\Omega \frac{t}{2}}
\Big[
\cosh{(|\xi_q^a|t)}\beta_q^a(0)
+\frac{\xi_q^a}{|\xi_q^a|}\sinh{(|\xi_q^a|t)}\beta_{-q}^{a\dagger}(0)
\Big]
\end{align}
for $|\omega_q^a-\Omega/2|\ll |\xi_q^a|$, and
\begin{equation}
\beta_q^a(t)\simeq
\textrm{e}^{-\textrm{i}\omega_q^a t}\beta_q^a(0) 
\end{equation}
for $|\omega_q^a-\Omega/2|\gg |\xi_q^a|$.
\end{subequations}

%===========================================================================
%===========================================================================
%===========================================================================
%===========================================================================
\section{Fermionic momentum distribution}
The results above prove the possibility of amplifying bosonic collective modes (with different wavenumbers for charge and spin modes) while parametrically modulating the underlying 1D fermionic many-body Hamiltonian. 
However, the detection of this amplification is not necessarily easy from the
experimental point of view. In the case of cold atoms in optical lattices this
would require a (spin-resolved) measurement of the cloud density during the
parametric modulation, without opening the trap. 
Very recently, in-situ measurements on confined ultracold atomic gases have been reported 
\cite{engel07_PRL,ott08_NAT,greiner09_NAT}. The spatial resolution of these measurements would allow for 
the detection of density modulations in the cloud associated to the parametric amplification of charge density waves. 
The detection of spin--charge separation, in addition to the rich physics of the Hubbard model for cold atoms in optical lattices, could motivate further experimental efforts towards the realization of local spin-resolved in-situ measurements not reported so far.

In view of this experimental challenge we now discuss the consequences of our analysis on the fermionic momentum distribution, which is the standard quantity measured in time-of-flight experiments \cite{bloch08_RMP}.
Due to the different amplified momenta for charge and spin modes, we show that the fermionic momentum distribution shows clear signatures of spin--charge separation.

The momentum distribution function
(per spin channel) for $\tau$-movers is defined as
$n_{k\tau}(t)=\int\textrm{d}x\,\textrm{e}^{\textrm{i}kx}
\left<\psi_{\tau\sigma}^\dagger(x, t)\psi_{\tau\sigma}(0, t)\right>$
where, within our perturbative treatment in $\gamma\ll1$, $\left<\cdots\right>$ represents a thermal average with respect to the
\textit{time-independent} part of the Hamiltonian \eqref{eq:H(t)}, \textit{i.e.}, for
$\gamma=0$. The fermionic 
field operators creating a $\tau$-mover with spin $\sigma$ at position $x$ and 
time $t$ can be expressed in terms of the bosonic operators as \cite{giuliani}
\begin{equation}
\label{eq:field}
\psi_{\tau\sigma}^\dagger(x, t)=
\frac{\textrm{e}^{-\textrm{i}\tau k_\textrm{F}x-\theta_{\tau\sigma}(x,
t)}}{\sqrt{L(1-\textrm{e}^{-2\pi/L\Lambda})}}U_{\tau\sigma}(t), 
\end{equation}
with $\theta_{\tau\sigma}(x, t)=[\theta_\tau^\textrm{c}(x,
t)+\sigma\theta_\tau^\textrm{s}(x, t)]/\sqrt{2}$ ($\sigma=1$ and $-1$ 
correspond to spin up and down, respectively) and 
$\theta_\tau^a(x, t)=\sum_{\tau q>0}(2\pi/L|q|)^{1/2}
\left[\textrm{e}^{\textrm{i}qx}b_q^a(t)-\textrm{e}^{-\textrm{i}qx}b_q^{a\dagger}(t)\right]$.
In \eqref{eq:field}, the
unitary Klein operator $U_{\tau\sigma}$ increases the number of $\tau$-movers with spin
$\sigma$ by one.

With \eqref{eq:field}, we find after a lengthy but straightforward
calculation
\begin{equation}
\left<\psi_{\tau\sigma}^\dagger(x, t)\psi_{\tau\sigma}(0, t)\right>=
\frac{\textrm{e}^{-\textrm{i}\tau k_\textrm{F}x+\phi_\tau(x,
t)}}{L(1-\textrm{e}^{-2\pi/L\Lambda})}.
\end{equation}
Here, within the rotating wave approximation, we have 
\begin{align}
\label{eq:phase}
\phi_\tau(x, t)=&\sum_{a, q>0}\frac{\pi}{Lq}
\Big\{
\textrm{i}\tau\sin{(qx)}\left[\beta_q^a(t), \beta_q^{a\dagger}(t)\right]
\nonumber\\
&+\left[\cos{(qx)}-1\right]\cosh{(2\varphi_q^a)}
\left<\left\{\beta_q^{a}(t), \beta_q^{a\dagger}(t)\right\}\right>
\Big\},
\end{align}
with $\beta_q^a(t)$ given in eq.~\eqref{eq:beta}.
The summation over $q$ in \eqref{eq:phase} is simplified by taking the evolution 
of the Bogoliubov operators in eq.~(\ref{eq:betaONOFF}) for wavenumbers belonging 
to the off-resonance and the narrow on-resonance windows. The latter are centered 
around the two resonant wavenumbers $q_\textrm{c/s}=\Omega /2v^\textrm{c/s}$ for 
the charge and spin modes, respectively. 

\begin{figure}[t]
\onefigure[width=\columnwidth]{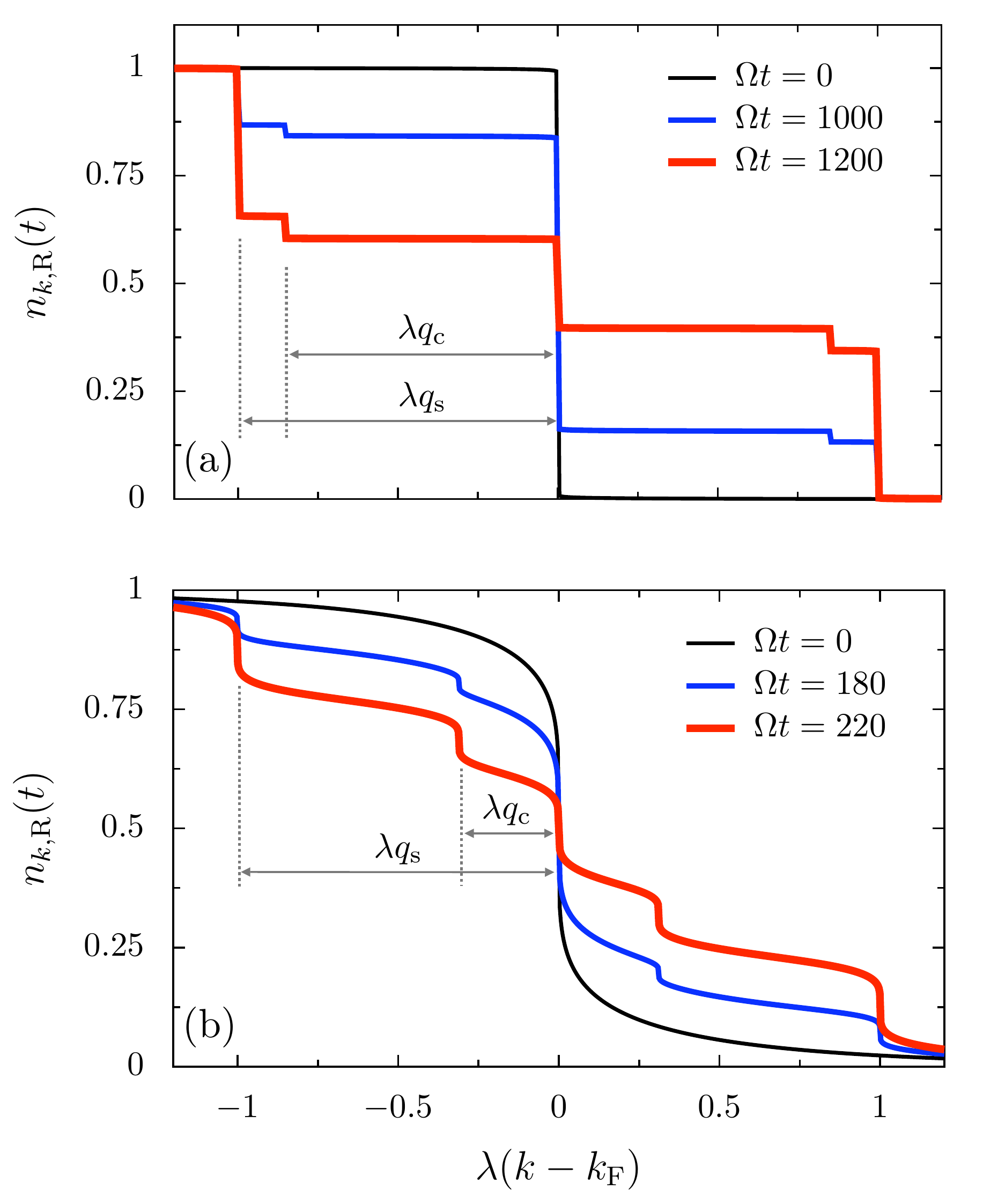}
\caption{\label{fig:momentum_distribution}%
Fermionic momentum distribution for right movers $n_{k,\textrm{R}}(t)$ in the
short-time regime
(see eq.~\ref{eq:nSMALL}) as a function of
momentum $k$, scaled by the screening length $\lambda$. 
In the figure, $T=0$, 
$\gamma=0.1$ and $\lambda q_\textrm{s}=1$.
(a) Case of contact $s$-wave scattering interaction, with
$V_0^\perp/v_\textrm{F}=0.5$ and $V_0^\parallel=V_{2k_\textrm{F}}^\parallel=0$, 
such that $\lambda q_\textrm{c}=\lambda q_\textrm{s}v_\textrm{s}
/v_\textrm{c}\simeq0.85$
(b) Case of finite-range interactions, with
$V_0^\parallel/v_\textrm{F}=V_0^\perp/v_\textrm{F}=10$ and
$V_{2k_\textrm{F}}^\parallel/v_\textrm{F}=2$,
such that $\lambda q_\textrm{c}=\lambda q_\textrm{s}v_\textrm{s}
/v_\textrm{c}\simeq0.31$. Thicker lines correspond to larger times $t$.}
\end{figure}

Our procedure allows for a fully analytical treatment of the problem, leading to
the fermionic correlator
\begin{equation}
\label{eq:correlator}
\left<\psi_{\tau\sigma}^\dagger(x, t)\psi_{\tau\sigma}(0, t)\right>=
\left<\psi_{\tau\sigma}^\dagger(x)\psi_{\tau\sigma}(0)\right>_0\mathcal{A}(x, t), 
\end{equation}
where the correlator without parametric amplification (\textit{i.e.}, for $\gamma=0$) 
in the zero temperature limit is \cite{giuliani}
\begin{equation}
\label{eq:correlator_0}
\left<\psi_{\tau\sigma}^\dagger(x)\psi_{\tau\sigma}(0)\right>_0=
\frac{\textrm{i}\textrm{e}^{-\textrm{i}\tau k_\textrm{F}x}}{2\pi(\tau x+\textrm{i}0^+)}
\left(\frac{\lambda^2}{x^2+\lambda^2}\right)^\alpha
\end{equation}
with
$\alpha=(\sinh^2{\varphi^\textrm{c}}+\sinh^2{\varphi^\textrm{s}})/2$. To obtain
\eqref{eq:correlator_0}, we
approximated $\varphi^a_q$ by its $q\rightarrow 0$ limit $\varphi^a$. This is justified
provided the sum over momenta $q$ in \eqref{eq:phase} is cutoff at $q\sim 1/\lambda$, where
$\lambda$ is the screening length associated with the specific form of the
interaction between the particles. This leads to the
fermionic momentum distribution in the absence of the parametric amplification
$n_{k, \tau}^0$. As the problem is fully symmetric between right and left movers, 
we focus on the former without loss of generality. 
For $k>k_\textrm{F}$, the momentum distribution can be expressed
as
\begin{equation}
n_{k, \textrm{R}}^0=\frac{2^{1/2-\alpha}}{\sqrt{\pi}\Gamma(\alpha)}
\int_0^\infty{\rm d}Q'(Q+Q')^{\alpha-1/2}\textrm{K}_{\alpha-1/2}(Q+Q'),
\end{equation}
with $Q=\lambda(k-k_\textrm{F})$, $\Gamma(z)$ and $\textrm{K}_\nu(z)$ being the
gamma and the modified Bessel functions, respectively. 
In particular, $n_{k-k_\textrm{F}, \textrm{R}}^0=1-n_{-k+k_\textrm{F}, \textrm{R}}^0$.
The function $n_{k, \textrm{R}}^0$ is presented in
fig.~\ref{fig:momentum_distribution} at time $t=0$ for contact and finite-range
interactions. The latter may be of relevance for the treatment of trapped cold
ions.

The amplification factor in eq.~(\ref{eq:correlator}) reads
\begin{equation}
\label{eq:A}
\mathcal{A}(x, t)=\exp{\left(\sum_{a=\textrm{c,s}}h^{a}_{}(t)\left[
\cos{(q_{a}x)}-1\right]\right)},
\end{equation}
where
\begin{equation}
\label{eq:h^a}
h^{a}_{}(t)=[1+2 n_\textrm{B}(\Omega/2)] \kappa^{a}_{}\cosh{(2\varphi^a)}
\left[\cosh{(\kappa^{a}\Omega t)}-1\right],
\end{equation}
and 
$\kappa^{a}=\gamma \left|
\sinh{(2\varphi^{a})}\right| v_\textrm{F}/2v^{a}$.
Here, $n_\textrm{B}(\omega)=(\textrm{e}^{\omega/T}-1)^{-1}$ is the Bose distribution at
temperature $T$. 
It is important to notice that the amplification factor equals 1 in the non-interacting limit (where $\sinh (\varphi^a)=0$). Once again, this highlights the importance of fermionic interactions as a necessary ingredient for the parametric amplification to occur.
Out of eq.~(\ref{eq:correlator}) the momentum distribution at finite times 
thus results in the convolution
\begin{equation}
\label{eq:nFIN}
n_{k,{\rm R}}(t)=\int_{-\infty}^{+\infty}\frac{\textrm{d}q}{2\pi}\, n_{k-q,{\rm R}}^{0}\tilde{\mathcal{A}}(q, t)
\end{equation}
with $\tilde{\mathcal{A}}(q, t)$ the Fourier transform of \eqref{eq:A}.

Two qualitatively different regimes occur in the small- or large-time regimes,
\textit{i.e.}, if $h^{a}_{}(t)\ll 1$ or $h^{a}_{}(t)\gg 1$ in eq.~\eqref{eq:A}. 
In the first case, valid up to times of order 
$t_0\simeq-\ln{([1+2n_\textrm{B}(\Omega/2)]\kappa^a\cosh{(2\varphi^a)})}/\kappa^a\Omega$, 
the expansion of \eqref{eq:A} yields
$\tilde{\mathcal{A}}(q, t)=2\pi\delta (q)
+\pi\sum_{a}h^{a}_{}(t)[\delta (q-q_{a}^{})+\delta
(q+q_{a}^{})-2\delta(q)]$, leading to  
\begin{equation}
\label{eq:nSMALL}
n_{k,{\rm R}}(t)=n_{k,{\rm R}}^{0}+\sum_{a}^{}h^{a}_{}(t)\left( 
\frac{n_{k+q_{a}^{},{\rm R}}^{0}+n_{k-q_{a}^{},{\rm R}}^{0}}{2}-n_{k,{\rm
R}}^{0}\right).
\end{equation}
Despite parametric amplification, the fermionic momentum distribution fulfills $n_{k-k_\textrm{F}, \textrm{R}}=1-n_{-k+k_\textrm{F}, \textrm{R}}$ as in the unperturbed case, guaranteeing particle-number conservation.   
For $k>k_{{\rm F}}^{}$ it shows two steps of size $h^{\textrm{c}/\textrm{s}}(t)/2$ 
involving fermionic states with momenta up to $q_{\textrm{c}/\textrm{s}}$ away from the 
Fermi level, as exemplified in fig.~\ref{fig:momentum_distribution}.

These are direct signatures of the parametric amplification of the charge and spin 
density waves and their observation can thus be used to detect spin--charge separation 
in interacting 1D Fermi systems. By spanning the external modulation frequency $\Omega$, the 
whole dispersion of the collective modes can be mapped. From the point of view of the 
measurement, our result is best visible in the ``short-time regime" where the expansion 
above holds, leading to two well resolved steps of size up to order
1/2. This fact is crucial in order to experimentally detect the amplification in
the momentum distribution against other smoothening factors, like,
\textit{e.g.}, trapping and finite temperatures. 

\begin{figure}[t]
\onefigure[width=.95\columnwidth]{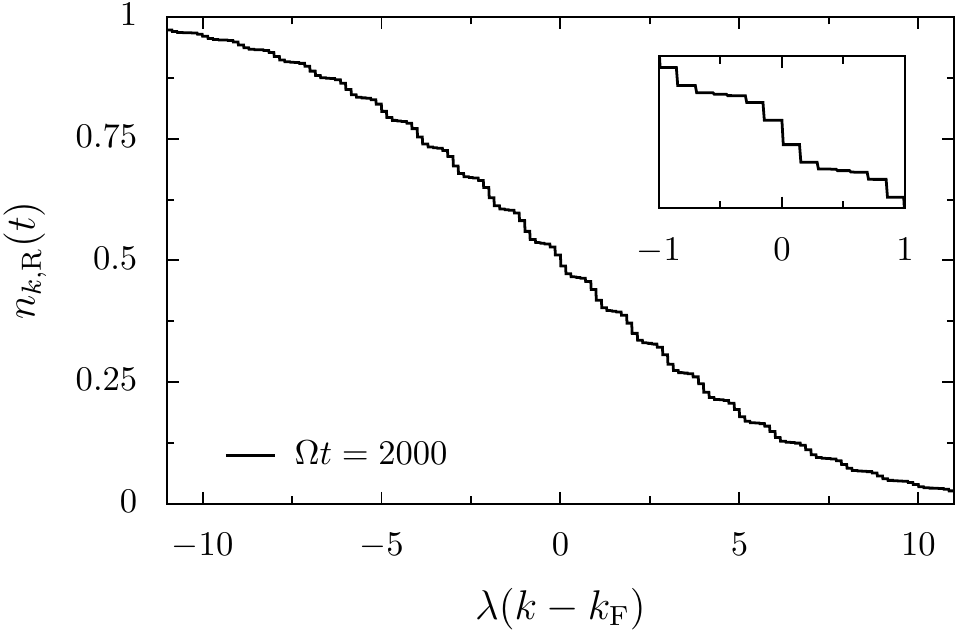}
\caption{\label{fig:longtime}%
Momentum distribution for right movers $n_{k,\textrm{R}}(t)$ in the long-time
regime (see eq.~\ref{eq:nLARGE}) as a function of momentum $k$. The parameters are the
same as in fig.~\ref{fig:momentum_distribution}a. The inset presents the
momentum distribution at a finer wavenumber scale close to the Fermi level.} 
\end{figure}

Our analytical treatment allows formally the analysis of the ``large-time regime" as well, 
where $h_{}^{a}(t)\gg 1$. In this case the amplification factor \eqref{eq:A} can be approximated as 
${\mathcal{A}}(x, t)=\mathcal{A}_{}^\textrm{c}(x, t)\mathcal{A}_{}^\textrm{s}(x, t)$, 
with $\mathcal{A}_{}^{a}(x,
t)=\sum_{n=-\infty}^{+\infty}\exp{\big(-(q_{a}x-2\pi n)^{2}h^{a}(t)/2\big)}$. 
As a consequence, the Fourier transform results in
\begin{equation}
\label{eq:AqLARGE}
\tilde{\mathcal{A}}(q, t)=\sum_{m,n=-\infty}^{\infty}f_{m}^\textrm{c}(t)f_{n}^\textrm{s}(t)\delta (q-mq_\textrm{c}^{}-nq_\textrm{s}^{})
\end{equation}
with $f_{m}^{a}(t)=\big(1/\sqrt{h^{a}(t)}\big)\exp{\big(-m^{2}_{}/2h_{}^{a}(t)\big)}$ leading to  
\begin{equation}
\label{eq:nLARGE}
n_{k,{\rm R}}(t)=\sum_{m,n=-\infty}^{\infty}\frac{f_{m}^\textrm{c}(t)f_{n}^\textrm{s}(t)}{2\pi }
n^{0}_{k-mq_\textrm{c}^{}-nq_\textrm{s}^{},{\rm R}}.
\end{equation}
Thus, in the large-time limit the fermionic momentum distribution shows many
small-size steps stemming
from both the charge and the spin sectors (see fig.~\ref{fig:longtime}). 
Indeed, the form of $f_{m}^{a}(t)$ shows 
how for large $h^{a}(t)$ more and more peaks of $\tilde{\mathcal{A}}(q, t)$ in
\eqref{eq:AqLARGE} become relevant. 
This will limit the experimental resolution of the structures in $n_{k,{\rm
R}}(t)$ in contrast 
to the short timescales.
Moreover, at large times the exponential amplification of bosonic modes requires a treatment of their residual interaction beyond the Luttinger liquid model, associated to parabolic corrections to the linearized spectrum around the 
Fermi level \cite{kagan09_PRA, piroo07_EPJB}. 
These effects lead to damping of the collective modes, which becomes relevant after a typical timescale $t_{\textrm{damp}}$. 
For $s$-wave scattering in the weak coupling limit $\eta=V_0^\perp/2\pi v_\textrm{F}\ll1$, $t_{\textrm{damp}}$ 
has been estimated to be  \cite{piroo07_EPJB}
$t_\textrm{damp}\simeq512E_\textrm{F}^2/\pi(\Omega\eta)^3$, 
with $E_\textrm{F}$ the Fermi energy. 
Our approximation of
neglecting such corrections is thus valid for pumping times $t\lesssim
t_\textrm{damp}$, while the spin--charge separation is best detectable in the ``short-time
limit" $t<t_0$. 
We have 
${t_0}/{t_\textrm{damp}}=\pi({\Omega}\eta/{E_\textrm{F}})^2
\ln{(2/\gamma\eta)}/256\gamma$.
For the parameters of fig.~\ref{fig:momentum_distribution}a, $t_0/t_\textrm{damp}\simeq0.004$. 
The regime of optimal visibility of the shoulders in the momentum distribution is thus well 
described by our ``non-interacting" collective modes approximation. Indeed, this picture is 
suitable to describe the ``long-time regime" $t_{0}^{}<t<t_\textrm{damp}$ of fig.~\ref{fig:longtime} as well, 
before damping yields saturation of the modes occupation at $t>t_\textrm{damp}$ \cite{kagan09_PRA}.

A final issue to be addressed in view of the experimental realization of our
proposal is the role of finite temperatures, where our analysis above applies as
well. The only differences are: (i) The presence of a non-vanishing Bose
distribution in eq.~\eqref{eq:h^a}. This yields a thermal seed for the amplification 
on top of the pure quantum fluctuations at $T=0$, and decreases the time needed for the formation of well-resolved steps in $n_{k,\tau}(t)$. 
(ii) The unperturbed momentum distribution $n_{k,{\rm R}}^{0}$ in
eq.~(\ref{eq:nFIN}) has to 
be replaced
with the finite temperature one, involving a thermal smearing of order $T$
around the Fermi level (for $\alpha\ll 1$) on top of that purely induced by interactions. 
As thermal smearing involves wavenumbers up to order $T/v_\textrm{F}$ around
$k_\textrm{F}$,
the shoulders in the final momentum distribution at short times are thus better visible 
if $v_{{\rm F}}q_{a}\simeq\Omega/2\gtrsim T$ ($a=\textrm{c}$, s), which can be 
guaranteed by choosing a sufficiently large $\Omega$ at a given temperature. 
In order for the Luttinger treatment to be reliable, the amplified $q_{a}$ 
should however be smaller than $k_{{\rm F}}^{}$, which restricts the best choice of 
$\Omega$ to the window $2T\lesssim \Omega\lesssim 4E_{{\rm F}}^{}$. The current
experimental efforts to reach regimes of very low-temperatures $T\ll T_{{\rm
F}}^{}$ with cold fermionic gases would then further improve the frequency range
for the best visibility of the spin--charge separation.

%===========================================================================
%===========================================================================
%===========================================================================
%===========================================================================
\section{Experimental realization}
Our proposal could be experimentally realized with an equal mixture of quantum
degenerate fermionic $^{40}$K atoms confined into 1D cigars in the two hyperfine states 
$|F, m_F\rangle=|9/2, -9/2\rangle=|\downarrow\rangle$ and 
$|F, m_F\rangle=|9/2, -7/2\rangle=|\uparrow\rangle$. Here, $F$ is the total
angular momentum and $m_F$ its projection along the quantization axis. It is important to realize
that the atomic density corresponds to the charge channel, while the two
hyperfine states above correspond to the (pseudo-)spin channel. Assuming the
cigar of length $L=0.1\,\textrm{mm}$ to be homogeneous and containing $N=10^2$ atoms,
we have $k_\textrm{F}=3\times 10^6\,\textrm{m}^{-1}$, which corresponds to a
Fermi energy and temperature of order $E_\textrm{F}/\hbar=7\,\textrm{kHz}$
and $T_\textrm{F}=60\,\textrm{nK}$, respectively. 
For neutral atoms, we assume contact $s$-wave scattering for which 
$V_0^\parallel=V_{2k_\textrm{F}}^\parallel=0$ as required by Pauli principle,
and $V_0^\perp/v_\textrm{F}=\pi k_{\textrm{F}}a\omega_\perp/E_\textrm{F}$ \cite{olsha98_PRL}. 
Here, $a$ is the 3D $s$-wave scattering length and $\omega_\perp$ the frequency
of the transverse confining lasers creating the cigars. 
Notice that $E_\textrm{F}\ll \omega_\perp$ justifies the effective 1D treatment
of the cigars. 
Assuming a scattering length of the order of
$a=10\,\textrm{nm}$ and $\omega_\perp=40\,\textrm{kHz}$, we obtain
$V_0^\perp/v_\textrm{F}=0.5$, which corresponds to the parameters in
fig.~\ref{fig:momentum_distribution}a.

As our proposal is optimized in
the regime $T\lesssim\Omega/2\lesssim 2E_\textrm{F}$, and assuming
$T/T_\textrm{F}\simeq0.2$, this corresponds to pumping
frequencies in the range $3\,\textrm{kHz}\lesssim\Omega\lesssim28\,\textrm{kHz}$.
Measuring the fermionic momentum distribution by a time-of-flight
experiment \cite{bloch08_RMP}, one should thus obtain 
a clear signature of 
spin--charge separation
for pumping
times of the order of $100\,\textrm{ms}$ (for $\Omega=10\,\textrm{kHz}$ and
$\gamma=0.1$), as exemplified in
fig.~\ref{fig:momentum_distribution}a. Notice that pumping for larger times
would lead to a situation similar to the one depicted in
fig.~\ref{fig:longtime} where spin--charge separation is
much less clearcut and where temperature effects are likely to smear out most
signatures of shoulders.

%===========================================================================
%===========================================================================
%===========================================================================
%===========================================================================
\section{Conclusion}

In this work we have shown the possibility of parametrically amplifying
collective modes in a modulated 1D fermionic many-body system. The amplification
is crucially affected by fermionic interactions which are here exactly treated
within the Luttinger liquid picture. This opens the perspective of similar
observations in systems of higher dimensionality as well.

Our analysis shows that the amplification of charge and spin density waves of
the Luttinger liquid results in clear steps in the fermionic momentum
distribution. The wavenumber extension of the steps directly reveals the
different momenta of the excited charge and spin modes and thus offers a tool
for the detection of spin--charge separation. In parallel, we show that
the best resolution of the steps is achieved by modulations of relatively short
times and that they survive thermal effects for large enough modulation
frequencies.

Our proposal of detection of spin--charge separation is particularly suitable
for systems of cold fermionic atoms in 1D optical lattices with modulated
intensity. The fermionic momentum distribution is indeed the standard quantity
measured in time-of-flight experiments. We stress that for our proposal no
additional experimental setup is required on top of the already present tunable
lasers creating the optical lattice.

%===========================================================================
%===========================================================================
%===========================================================================
%===========================================================================
\acknowledgments
We benefited from fruitful discussions with \textsc{T.\ Brandes, T.\ Esslinger, 
T.\ Giamarchi, L.\ Goren,} and \textsc{A.\ Stern}. CDG enjoyed the hospitality of the Indian
Institute of Technology Roorkee while this work was completed.\\

\noindent\textit{Additional remark:}
During the completion of this
work, we became aware of ref.~\cite{kagan09_PRA} where similar effects have been
investigated.


\begin{thebibliography}{}

\bibitem{landau}
\Name{Landau L. D. \and Lifshitz E. M.}
\Book{Mechanics} 
\Publ{Pergamon Press}
\Year{1976}.

\bibitem{bloch08_RMP}
\Name{Bloch I., Dalibard J. \and Zwerger W.} 
\REVIEW{Rev. Mod. Phys.}{80}{2008}{885}.

% parametric resonance (experiments)
\bibitem{stofe04_PRL}
\Name{St\"oferle T. \textit{et al.}} %, Moritz H., Schori C., K\"ohl M. \and Esslinger T.} 
\REVIEW{Phys. Rev. Lett.}{92}{2004}{130403}.

\bibitem{engel07_PRL}
\Name{Engels P., Atherton C. \and Hoefer M. A.}
\REVIEW{Phys. Rev. Lett.}{98}{2007}{095301}.

\bibitem{tozzo05_PRA}
\Name{Tozzo C., Kr\"amer M. \and F.\ Dalfovo} 
\REVIEW{Phys. Rev. A}{72}{2005}{023613}.

\bibitem{goren07_PRA}
\Name{Goren L., Mariani E. \and Stern A.} 
\REVIEW{Phys. Rev. A}{75}{2007}{063612}.

\bibitem{giuliani}
\Name{Giuliani G. F. \and Vignale G.}
\Book{Quantum Theory of the Electron Liquid} 
\Publ{Cambridge University Press}
\Year{2005}.

\bibitem{giamarchi}
\Name{Giamarchi T.}
\Book{Quantum Physics in One Dimension} 
\Publ{Oxford University Press}
\Year{2004}.

% spin-charge probed w/ transport
\bibitem{ausla05_Science}
\Name{Auslaender O. M. \textit{et al.}} % Steinberg H., Yacoby A., Tserkovnyak Y., Halperin B. I., Baldwin K. W., 
%Pfeiffer L. N. \and West K. W.} 
\REVIEW{Science}{308}{2005}{88}.

% exp. spin-charge ARPES
\bibitem{kim06_NaturePhysics}
\Name{Kim B. J. \textit{et al.}} %, Koh H., Rottenberg E., Oh S.-J., Eisaki H., Motoyama N., Uchida
%S., Tohyama T., Maekawa S., Shen Z.-X. \and Kim C.}
\REVIEW{Nature Phys.}{2}{2006}{397}.

\bibitem{recat03_PRL}
\Name{Recati A. \textit{et al.}}
\REVIEW{Phys. Rev. Lett.}{90}{2003}{020401}.

\bibitem{kecke05_PRL}
\Name{Kecke L., Grabert H. \and H\"ausler W.}
\REVIEW{Phys. Rev. Lett.}{94}{2005}{176802}.

\bibitem{kolla05_PRL}
\Name{Kollath C., Schollw\"ock U. \and Zwerger W.}
\REVIEW{Phys. Rev. Lett.}{95}{2005}{176401}.

\bibitem{polin07_PRL}
\Name{Polini M. \and Vignale G.}
\REVIEW{Phys. Rev. Lett.}{98}{2007}{266403}.

\bibitem{klein08_PRA}
\Name{Kleine A. \textit{et al.}} %, Kollath C., McCulloch I. P., Giamarchi T. \and Schollw\"ock U.}
\REVIEW{Phys. Rev. A}{77}{2008}{013607}.

\bibitem{mathe08_PRL}
\Name{Mathey L., Altman E. \and Vishwanath A.}
\REVIEW{Phys. Rev. Lett.}{100}{2008}{240401}.

\bibitem{kagan09_PRA}
\Name{Kagan Yu. \and Manakova L. A.}
\REVIEW{Phys. Rev. A}{80}{2009}{023625}.

\bibitem{jorde08_Nature}
\Name{J\"ordens R. \textit{et al.}} %, Strohmaier N., G\"unter K., Moritz H.\ and Esslinger T.}
\REVIEW{Nature}{455}{2008}{204}.

\bibitem{schne08_Science}
\Name{Schneider U. \textit{et al.}} %, Hackerm\"uller L., Will S., Best Th., Bloch I., Costi T. A.,
%Helmes R. W., Rasch D. \and Rosch A.}
\REVIEW{Science}{322}{2008}{1520}.

\bibitem{masse09_PRL}
\Name{Massel F., Leskinen M. J. \and T\"orm\"a P.}
\REVIEW{Phys. Rev. Lett.}{103}{2009}{066404}.

\bibitem{iucci06_PRA}
\Name{Iucci A. \textit{et al.}}
\REVIEW{Phys. Rev. A}{73}{2006}{041608(R)}.

\bibitem{ott08_NAT}
\Name{Gericke T. \textit{et al.}} 
\REVIEW{Nature Phys.}{4}{2008}{949}.

\bibitem{greiner09_NAT}
\Name{Bakr W. S. \textit{et al.}} 
\REVIEW{Nature}{462}{2009}{74}.

\bibitem{piroo07_EPJB}
\Name{Pirooznia P. \and Kopietz P.}
\REVIEW{Eur. Phys. J. B}{58}{2007}{291}.

\bibitem{olsha98_PRL}
\Name{Olshanii M.}
\REVIEW{Phys. Rev. Lett.}{81}{1998}{938}.

\end{thebibliography}
\end{document}